\documentclass[
    ,final            % use final for the camera ready runs
  ]
  {aipproc}

\layoutstyle{8x11double}
\newcommand{\apj}{ApJ}
\newcommand{\apjl}{ApJL}
\newcommand{\mnras}{MNRAS}
\newcommand{\aap}{A\&A}

\begin{document}

\title{Theory of Pulsar Wind Nebulae}


\author{Bucciantini, N.}{
  address={Astronomy Department, University of California at Berkeley, 601 Camblell Hall, Berkeley, CA, 947
06, USA}
}

\begin{abstract}
Our understanding of Pulsar Wind Nebulae (PWNe), has greatly improved 
in the last years thanks to unprecedented high resolution images taken 
from the HUBBLE, CHANDRA and XMM satellites. The discovery of complex but
similar inner features, with the presence of unexpected axisymmetric rings
and jets, has prompted a new investigation into the dynamics 
of the interaction of the pulsar winds with the surrounding SNR, which, 
thanks to the improvement in the computational resources, has let to a  
better understanding of the properties of these objects. On the other  
hand the discovery of non-thermal emission from bow shock PWNe, and   
of systems with a complex interaction between pulsar and 
SNR, has led to the development of more reliable evolutionary models. I 
will review the standard theory of PWNe, their evolution, and the current 
status in the modeling of their emission properties, in particular I 
will show that our evolutionary models are able to describe the observations, 
and that the X-ray emission can now be reproduced with sufficient accuracy, 
to the point that we can use these nebulae to investigate fundamental issues 
as the properties of relativistic outflows and particle acceleration. 

\end{abstract}

%%%%%%%%%%%%%%%%%%%%%%%%%%%%%%%%%%%%%%%%%%%%%%%%%%%%%%%%%%%%%%%%%%%
%%
%% The below \maketitle command inserts the actual front matter data.
%% It has to follow the above declarations.
%%
%%%%%%%%%%%%%%%%%%%%%%%%%%%

\maketitle

%%%%%%%%%%%%%%%%%%%%%%%%%%%%%%%%%%%%%%%%%%%%
%% MAINMATTER
%%
%%%%%%%%%%%%%%%%%%%%%%%%%%%%%%%%%%%%%%%%%%%%%%%%%%%%%%%%%%%%%%%%%%%%%%%%%%%%
%% Headings:
%%
%% The aipproc supports three heading levels, i.e., \section,
%%	\subsection, and \subsubsection.
%%%%%%%%%%%%%%%%%%%%%%%%%%%%%%%%%%%%%%%%%%%%%%%%%%%%%%%%%%%%%%%%%%%%%%%%%%%%
%% Cross-references:
%%
%% Page numbers (\pageref) and headings can NOT be referenced in the class,
%% since before being produced, no page numbers are determined.
%%
%% Tables, figures, and equeations can be referenced by using the LaTex
%% 	commands \label and \ref. For references to equation numbers, \eqref
%%	can be used, which will print "(1)" (while \ref will result in "1").
%%
%%%%%%%%%%%%%%%%%%%%%%%%%%%%%%%%%%%%%%%%%%%%%%%%%%%%%%%%%%%%%%%%%%%%%%%%%%%%
%% Lists: 
%%
%% Standard "itemize", "enumerate", etc. list environments are supported.
%%%%%%%%%%%%%%%%%%%%%%%%%%%%%%%%%%%%%%%%%%%%%%%%%%%%%%%%%%%%%%%%%%%%%%%%%%%%
%% Urls:
%%
%% \url{} command is provided for documenting URLs.
%%%%%%%%%%%%%%%%%%%%%%%%%%%%%%%%%%%%%%%%%%%%

\section{Introduction}

Pulsar Wind Nebulae (PWNe) are bubbles of relativistic particles and magnetic field created when the ultra-relativistic wind from a pulsar interacts with the ambient medium, either SNR or ISM. The best example of a PWN, often considered the prototype of this entire class of objects, is the Crab Nebula. The first theoretical model of PWNe was presented by \citet{ree74}, developed in more details by  \citet{ken84a,ken84b} (KC84 hereafter), and is based on a relativistic MHD description. The ultra-relativistic pulsar wind is confined inside the SNR, and slowed down to non relativistic speeds in a strong termination shock (TS). At the shock the toroidal magnetic field of the wind is compressed, the plasma is heated and particles are accelerated to high energies. A bubble of high energy particles and magnetic field is produced where the post-shock flow expands at a non relativistic speeds toward the edge of the nebula.

Despite its simplicity the MHD model can explain many of the observed properties of PWNe. Acceleration at the TS accounts for the continuous, non-thermal, very broad-band spectrum, extending from Radio to X-rays, with spectral index in the range 0-1.2, steepening with increasing frequencies, and modeled as synchrotron emission \citep{ver93,ban99,wei00,wil01,mor04}. The under-luminous region, centered on the location of the pulsar, is interpreted as the ultra-relativistic unshocked wind. Polarization measures \citep{wil72,vel85,sch79,hic90,mic91} show that emission is highly polarized and the nebular magnetic field is mostly toroidal, as one would expect from the compression of the pulsar wind. This model also predicts that PWNe should appear bigger at smaller frequencies: high energy X-rays emitting particles have a short lifetime for synchrotron losses, and they are are present only in the vicinity of the TS; in contrast the synchrotron lifetime for Radio particles is longer than the age of the nebula, so they fill the entire volume. This increase in size at smaller frequencies is observed in the Crab Nebula \citep{ver93,bie97,ban98}. If one considers the pressure anisotropy due to the compressed nebular toroidal magnetic field  \citep{beg92,van03}, it is also possible to recover the elongated axisymmetric shape of many PWNe ({\it i.e.} Crab Nebula, 3C58).

By comparing observations with the predictions of the simple spherically symmetric MHD model it is possible to  constraint some of the properties of the pulsar wind, at least at the distance of the TS. To explain the dynamics of the plasma, as well as the emission properties of the nebula, the Lorentz factor of the wind is estimated to be $\sim 10^6$, and the ratio between Poynting flux and kinetic energy $\sigma\sim 0.003$. This shows that nebular properties can be used to derive informations on the conditions of the pulsar wind at large distances.

\section{Evolution of PWNe}

In the analytic model developed by KC84 the SNR has only a passive role, providing the confinement of the PWN. It is however known that the details of the flow structure in the PWN and its emission properties depend critically on the boundary condition with the SNR. Given the complexity of the interaction and the existences of different evolutionary phases, a detailed study of the evolution of the system PWN-SNR, has been possible only recently, thanks to the improvement in computational resources \citep{van01,blo01,me03,van04}. 

%%%%%%%%%%%%%%%%%%%%%%%%%%%%%%%%%%% FIG 1 %%%%%%%%%%%%%%%%%%
\begin{figure}
\label{fig:1}
\resizebox{7.8cm}{!}{
\includegraphics[scale=0.7]{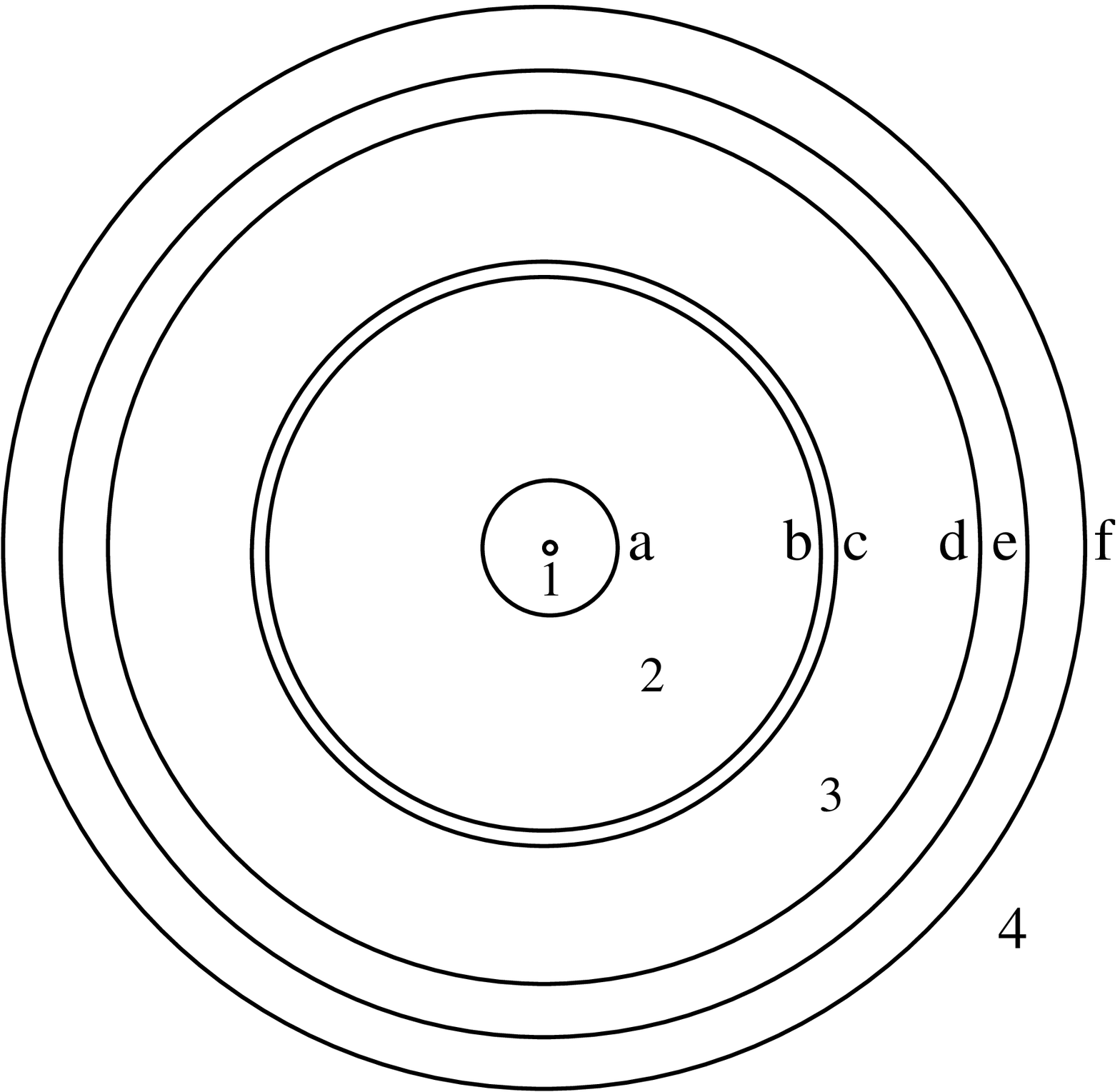}\hspace{0.5truecm}\includegraphics{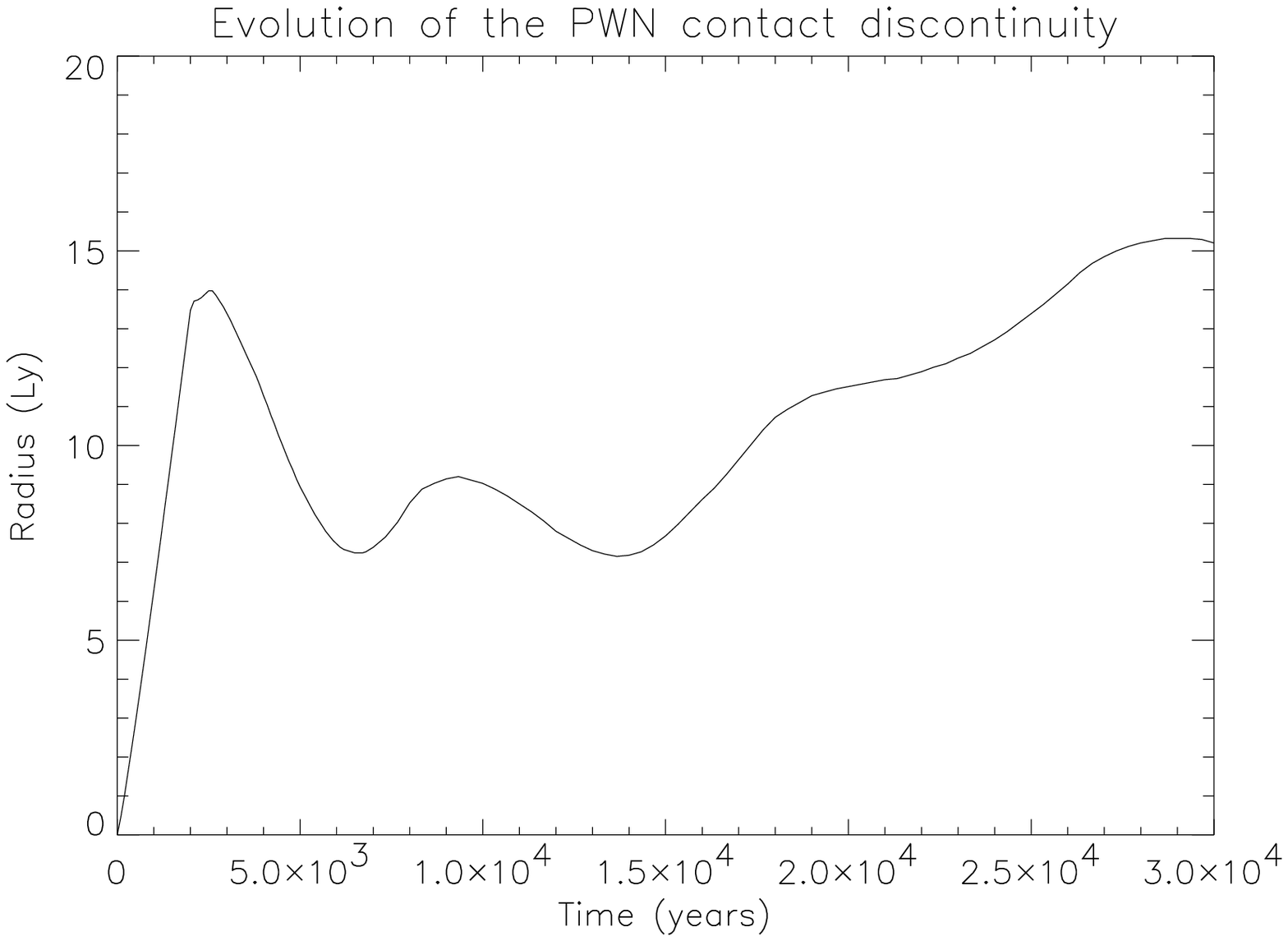}}
\caption{Left picture: schematic representation of the  global structure of the PWN in the first phase of its evolution inside a SNR.  From the center the various regions are: 1- the  relativistic pulsar wind, 2- the hot magnetized bubble responsible for the non thermal emission, 3- the free expanding ejecta of the SNR, 4- the ISM. These regions are separated by discontinuities: a- the wind termination shock, b- the contact discontinuity between the hot shocked pulsar material and the swept-up SNR ejecta, c-the front shock of the thin shell expanding into the ejecta, d- the reverse shock of the SNR, e- the contact discontinuity separating the ejecta material from the compressed ISM, f- the forward SNR shock. Right picture: evolution of the PWN size, from free-expansion to sedov phase (from \citet{me03}).}
\end{figure}
%%%%%%%%%%%%%%%%%%%%%%%%%%%%%%%%%%%%%%%%%%%%%%%%%%%%%%%%%%%%

It is immediately evident by comparing the energy in the SNR ($\sim 10^{51}$ ergs) to the total energy injected by the pulsar during its lifetime ($\sim 10^{49}$ ergs) that while the PWN cannot significantly affect the SNR, the evolution of the SNR can have important consequences for the PWN.

A simple 1D model for PNWe evolution shows the existence of three main phases (for a more complete discussion of PWN-SNR evolution see \citet{ren84} and \citet{gae06}). At the beginning the PWN expands inside the cold SN ejecta. The SN ejecta are in free expansion, so this phase is somehow referred as {\it free expansion phase}. During this phase, that lasts for about 1000-3000 yr, the pulsar luminosity is high and almost constant. This is the present phase of the Crab Nebula, and PWNe in this phase are expected to shine in high energy X-rays emission.  The expansion velocity of PWNe in this early stage is known to be a few thousands kilometer per second, much higher than typical pulsar velocities in the range 50-300 km/s. For this reason one can neglect the pulsar kick in modeling young objects, and assume the pulsar to be centrally located.  As the system expands inside the high density, cold, supersonic ejecta of the SNR, a thin shell of compressed material is formed. The evolution of this shell can be easily described in the thin-shell approximation, and it is possible to derive an analytic self-similar solution \citep{me04b}. Both analytic and numerical results show that the thin shell accelerates with time. Given that the density of the shell is much higher than the enthalpy of the relativistic plasma, the shell is subject to Rayleigh-Taylor instability. This is supposed to be at the origin of the filamentary network of the Crab Nebula \citep{hes96,jun98,me04b}.

%%%%%%%%%%%%%%%%%%%%%%%%%%%%%%%%%%% FIG 2 %%%%%%%%%%%%%%%%%%
\begin{figure}
\label{fig:2}
\resizebox{16cm}{!}{\includegraphics[bb=0 224 305 546, clip, scale=2.6]{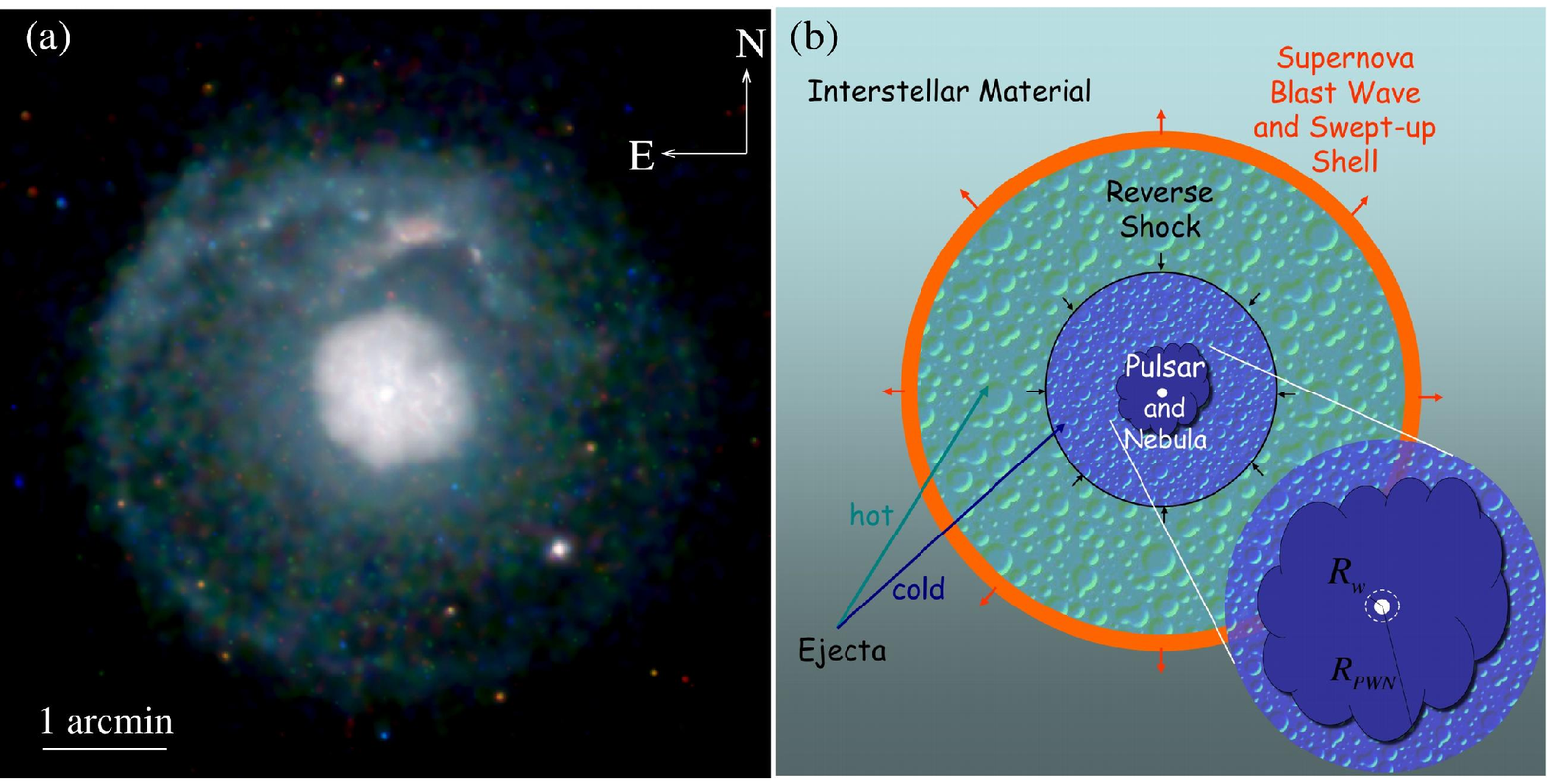}\hspace{0.5truecm}\includegraphics[angle=90,scale=2.2]{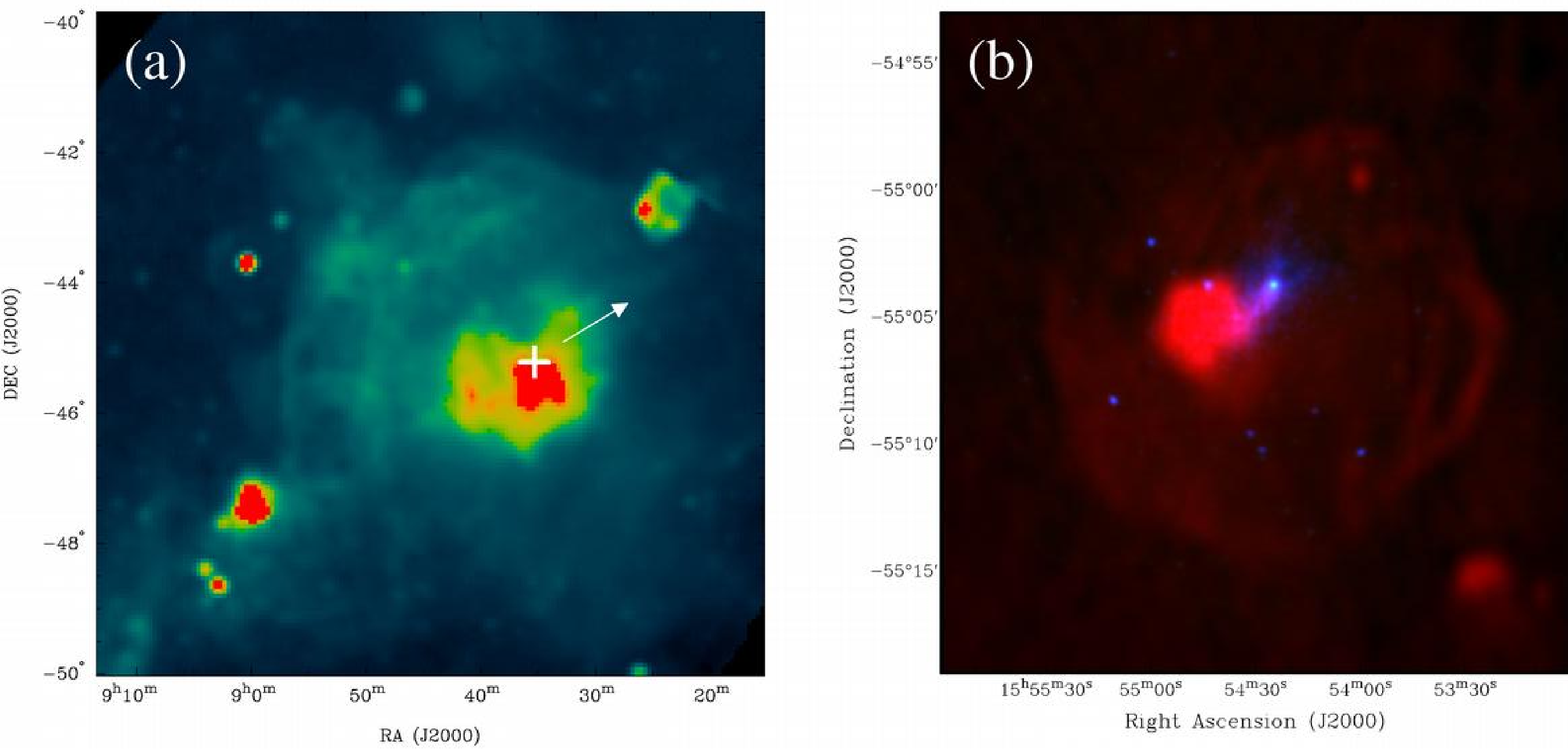}\hspace{0.5truecm}\includegraphics[scale=2.6]{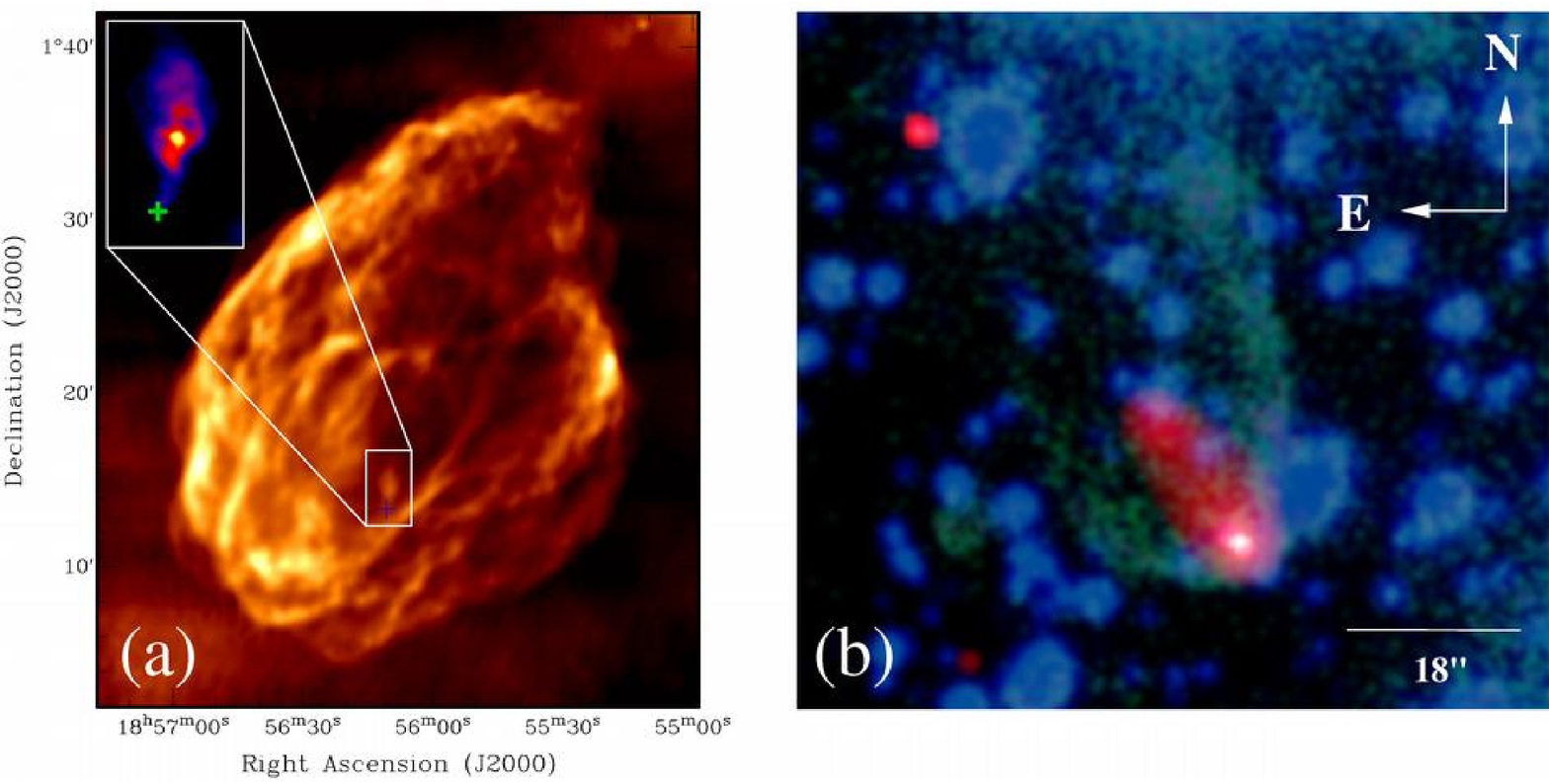}}
\end{figure}
\begin{figure}
\resizebox{16cm}{!}{\includegraphics[bb=378 52 565 748, clip, angle=90,scale =1.5]{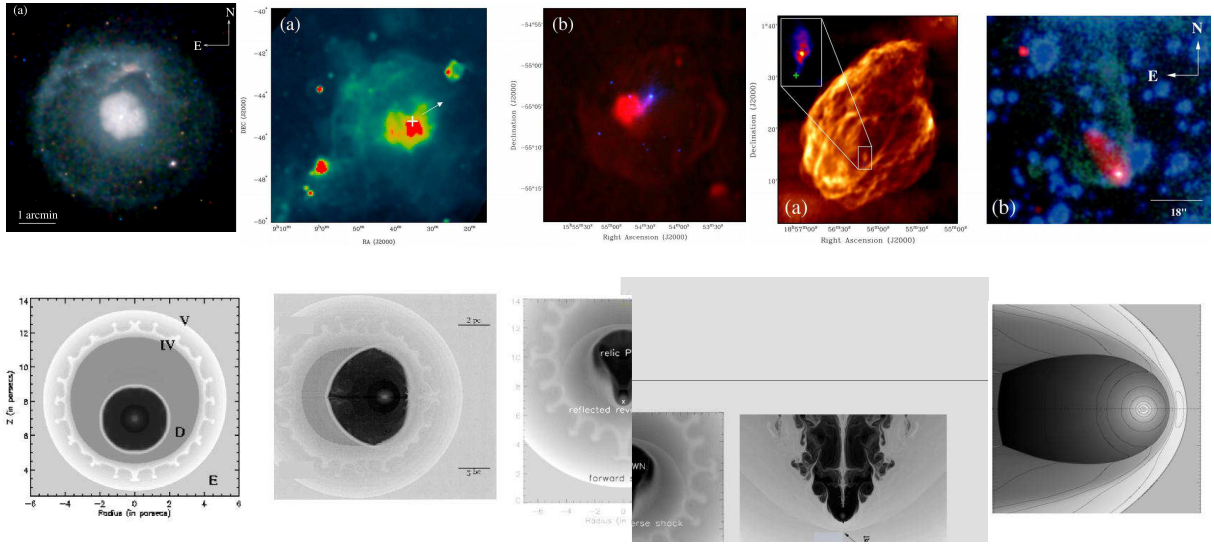}\includegraphics[bb=195 579 365 733, clip, angle=90,scale =1.5]{model.ps}}
\caption{Upper row: images of PWNe in various evolutionary phases (from Gaensler \& Slane \citep{gae06}. From Left to right: X-ray image of the composite remnant G21.5-0.9, free expansion phase; radio image of Vela SNR, displacement of the nebula due to the compression of the reverse shock during reverberation; SNR G 327.1-1.1 in radio (red) and X-rays (blue), relic PWN phase; W44 in radio, transition to the internal bow-shock phase; PSR B1957+20 in H$_\alpha$ (green) an X-rays (red), ISM bow-shock nebula. Numerical simulations of the various phases of the PWN-SNR interaction in the case of moving pulsars (from \citep{van04, me02}). Each figure of the second row corresponds to systems shown in the first one.}
\end{figure}
%%%%%%%%%%%%%%%%%%%%%%%%%%%%%%%%%%%%%%%%%%%%%%%%%%%%%%%%%%%%

During the first phase the PWN is inside the SNR shell. The expanding PWN will eventually reach the reverse shock in the SNR shell, which is supposed to recede to the center of the SNR in a time of order of 5000-10000 yr. From this moment on the evolution of the PWN is modified by the more massive and energetic SNR shell, and experiences a compression phase  generally referred as {\it reverberation phase}. In the simple 1D scenario the compressed PWN will then push back the ejecta, and the nebula might undergo several compressions and rarefactions. However more appropriate multidimensional studies have shown that the SNR-PWN interface is highly Rayleigh-Taylor unstable during compression \citep{blo01}, which can cause efficient mixing of the pulsar wind material with the SNR. This reverberation phase is supposed to last about $10^4-10^5$ years. Even if energy injection from the pulsar at these later times is negligible, PWNe can still be observed, due to the re-energization during compressions. The interaction with the reverse shock can lead to a variety of different morphological structures if one consider also the pulsar proper motion \citep{van03b,van04,me05}. Given that the evolution is now driven by the SNR reverse shock, the nebula can be displaced with respect to the location of the pulsar. At the beginning this might result into a system where the pulsar is not located at the center of the non thermal emission, analogous to what is observed in Vela. As the system evolves the reverse shock will completely displace the body of the PWN, creating a relic nebula. The relic PWN will mostly contain low energy particles, and will be visible in radio, while high energy particles, observable in X-rays will be seen only close to the pulsar. G24.5 shows indeed this kind of morphology \citep{gae06}. At later time the SNR starts cooling and eventually the pulsar will become supersonic with respect to the ejecta. Interestingly the location with respect to the SNR where this happen does not depend on the pulsar proper motion and turns out to be $\sim 70$\% of the radius of the forward shock. The pulsar will form around itself a bow-shock PWN, and one expects an emission tail to form connecting the pulsar to the relic PWN. This model apply to the morphology and structure of W44 \citep{gae06}.

The ultimate phase of a PWN evolution depends on the pulsar kick. For slow moving pulsars the PWN will expands adiabatically inside the heated SNR, now in Sedov phase. Given the absence of  energy injection, a PWN in this stage is probably only observable as a faint extended radio source. Fast moving pulsars can escape from the SNR. They will form bow-shock nebulae due to the interaction with the ISM, through which they are moving at supersonic speeds \citep{me01,me05}. These objects are observed both in $H_\alpha$ emission, due to ionization of ISM neutral hydrogen, and as long extended cometary-like source of non thermal emission, due to the shocked pulsar wind, now forced to flow in the direction opposite to the pulsar motion.  

\section{Inner structure}

\begin{figure}
\label{fig:3}
\resizebox{17cm}{!}{\includegraphics[bb=378 52 565 748, clip, angle=90,scale =1.5]{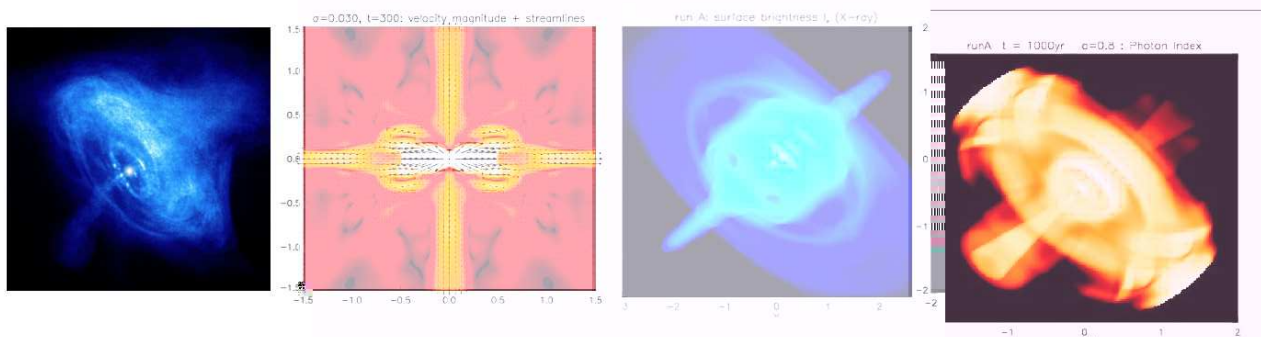}}
\caption{Fro Left to right: X-ray image of the crab nebula, jet-torus structure. Numerical result of the internal dynamics in a PWN, note the oblate shape of the TS, which focus the downstream flow toward the equator, \citep{ldz04}. The flow is then diverted back toward the axis but the magnetic hoop stresses, and is collimated into a jet (in this specific case there is an unmagnetized region around the equator, and an equatorial floe survives). Simulated X-rays synchrotron map based on numerical simulations and simulated map of the photon index from \citep{ldz06}. }
\end{figure}

Recent optical and X-rays images from HST, CHANDRA and XMM-Newton have triggered new interest of the scientific community for PWNe. The new data show that the inner region of PWNe is characterized by a complex axisymmetric structure, generally referred as {\it jet-torus structure} (Fig.~\ref{fig:3}). This was first observed in Crab \citep{hes95,wei00}, and has subsequently been detected in many other PWNe \citep{got00,gae01,hel01,pav03,gae02,lu02,rom03,sla04,cam04,rom05}. It is characterized by a main emission torus, corresponding to what is thought to be the equatorial plane of the pulsar rotation. Multiple arcs or rings are often present together with a central knot, located in the vicinity of the pulsar, and one or two opposite jets along the polar axis, which seems to originate close to the pulsar itself. Such structure cannot be explained in the simple 1D radial model by KC84. Even if a higher equatorial energy injection \citep{bog02a} can qualitatively account for the existence of a main torus it is not possible to reproduce the observed luminosity using the KC84 model. \citet{shi03} was the first to point that, the difference in brightness between the front and back sides of the torus in Crab Nebula, requires a post-shock flow velocity $\sim 0.3-0.4 c$, much higher than what expected for subsonic expanding flows. Moreover the existence of an inner ring, separated from the torus, contradicted the assumption of a smooth flow from the TS. Even more puzzling was the knot which seemed to be located inside the wind region. However, the most interesting feature was the jet \citep{lyu01}, because theoretical \citep{beg94,bes98} and numerical \citep{con99,bog01,gru05,kom06,me06} studies of relativistic winds from pulsars have shown no presence of collimated energetic outflow. Finally photon index maps of Crab Nebula \citep{mor04}, and now of Vela \citep{pav07}, show that the spectrum  flattens moving from the inner ring toward the main torus, when a steepening due to synchrotron losses is expected. Interestingly the symmetry axis of the jet-torus  appears to correspond to the major axis of the nebula, suggesting that the toroidal magnetic field is a key element in shaping the inner flow.

Only recently it was recognized the importance that magnetization and energy distribution in the pulsar wind have in determining the nebular flow. In contrast with the isotropic energy flux assumed in KC84, it was known for a long time that the asymptotic solution by \citet{mic73},  recently confirmed with numerical simulations \citep{bog01,kom06,me06}, predicts a higher equatorial flux, which naturally produces an oblate TS with a cusp in the polar region \citep{bog02a,bog02b}. Given that the obliquity of the TS at higher latitudes, the post shock flow in the nebula can have speeds $\sim 0.3-0.5 c$. Moreover if hoop-stresses are more efficient in the mildly relativistic flow, and the collimation of a jet might occur in the post shock region \citep{lyu02,kan03}. The evident complexity of this scenario clearly prevented any sophisticated theoretical model. Only recently, thanks to more efficient and robust  numerical schemes for relativistic MHD \citep{kom99,ldz03,gam03}, a detailed numerical description has been possible. Various numerical results show that, based on the theoretical assumption of axisymmetric energy flux and post-shock hoop stresses, it is possible to reproduce the observed features. This picture has been subsequently refined by investigating if and how the emission features could be used to understand the conditions of the pre-shock wind, trying to constrain many of the nebular emission properties, like polarization and photon index. More recently numerical studies have shown that also the short time variability might be recovered.

If one assume for the wind this solution of the force free model, the energy flux in the wind has a strong latitudinal dependence of the form $L(\theta)=L_o(1+\alpha\sin{(\theta)})$, where $\alpha$ is a measure of the pole-equator anisotropy, while the magnetic field in the wind $B(\theta)\propto \sin{(\theta)}$. Various numerical simulations of the interaction of such wind with the SNR ejecta have been presented \citep{kom04,ldz04,bog05,ldz06}. Interestingly these results show that the post shock flow is independent on the specific values of Lorentz factor or density distribution. In this sense flow dynamics cannot be used to constrain the value of the wind Lorentz factor or the multiplicity in the wind. In Fig.~\ref{fig:3} we can see the oblate shape of the TS.  Due to the TS shape, the flow is slowed down to speed $\sim c/3$ close to the equator, but at higher latitudes, where the shock is oblique, the post shock flow is still supersonic. The result of the anisotropic energy distribution in the wind is that almost all of the downstream plasma is funneled along the equatorial plane. The inner part  moves with speed $\sim c/3$, while the outer channels has velocity $\sim 0.5c$, the value expected in order to justify the luminosity distribution in the torus of the Crab Nebula \citep{shi03}.

%%%%%%%%%%%%%%%%%%%%%%%%%%%%%%%%%%%% FIG 3 %%%%%%%%%%%%%%%%%%
%%%%%%%%%%%%%%%%%%%%%%%%%%%%%%%%%%%%%%%%%%%%%%%%%%%%%%%%%%%%

As the flow expands in this equatorial sheet away from the TS, the magnetization increases, until equipartition is reached. Given the magnetic field distribution in the wind, the flow reaches equipartition close to the equator before that at higher latitudes. The post-shock flow is only mildly relativistic, so it can be efficiently collimated by hoop stresses. Once equipartition is reached, the magnetic pressure prevents further compression of the magnetic field, and the flow is diverted  back toward the axis. This is the process that causes collimation and the formation of a jet along the axis itself as seen in Fig.~3. The formation of the jet strongly depends on the values of the wind magnetization: if it is too small, then equipartition is not reached inside the nebula, the equatorial channel survives to the edge of the PWN and no jet is formed. At higher magnetization  equipartition is reached in the close vicinity of the TS, and most of the plasma is diverted and collimated in a jet. The size and flow velocity in the jet are a function of $\sigma$. For values $\sigma\sim 0.03$ the plasma speed in the jet is $\sim 0.7 c$ in agreement with observation of the jet in Crab Nebula \citep{wei00,hes02,mel05}. There is a global meridional circulation inside the nebula associated with the jet formation , with typical speeds $\sim 0.1 c$ that might lead to local shear instabilities and on the possible mixing with cold ions \citep{lyu03}.

It is also possible in numerical models to take into account modification of the wind properties associated with oblique rotators. While the energy distribution in the wind is identical to the aligned case \citep{bog01b,spi06}, the presence of a folded current sheet that might extend to higher latitudes, has important consequences on the magnetization. If this striped wind region is dissipated, and this can happen either in the wind \citep{lyu01b,kir03} or at the termination shock itself \citep{lyu03b,lyu05}, then the magnetization will vanish toward the equator and will reach a maximum values at intermediate latitudes, depending on the obliquity of the pulsar. This unmagnetized region close to the equator, adds complexity to the picture presented above. There is now an equatorial flow where equipartition is not reaches inside the nebula, that survives to much larger distances from the termination shock. However equipartition can be reached at higher latitudes and the flow is then diverted to the axis and collimated into a jet. Fig.~3 show the flow structure. Emission maps based on the results of numerical simulations suggest that a striped wind model is more promising in explaining the observes inner-ring outer-torus structure of many PWNe, while models without a neutral equatorial region seem more promising in the case of single ring nebulae. This show how emission features can be used as a probe for the properties of the wind. In particular recent results \citep{ldz06} have shown that, given the same average magnetization, a bigger striped wind region, leads to weaker jets, less intense hoop-stresses, and a bigger size of the termination shock.

Relativistic MHD simulations clearly show the formation of a jet, however to properly asses their validity and for a detailed comparison with observations it is necessary to compute the nebular emission. Despite several theoretical issues like the particles energy distribution at the injection and the efficiency of  acceleration, that can  modify the results, as a first approximation one might assume a uniform power-law injection distribution. Under this assumption, to obtain X-ray maps, one must then consider also the adiabatic and synchrotron losses, as fluid particles move in the nebula. In \citet{ldz06} the maximum energy $\epsilon_{max}$ of the particles' distribution is evolved with the flow, accounting both for synchrotron and adiabatic losses. In this case the particle distribution remains a power low with a break at $\epsilon_{max}$. A more sophisticated approach would require solving for the entire particles distribution at different energies.

In Fig.~3, we show a CHANDRA image of the Crab Nebula, and a map based on a simulation with striped wind in the X-rays. The knot and the inner ring are both explained as due to the high velocity flow in the immediate post shock region, at intermediate latitudes. The main torus observed in the Crab Nebula is due to the emission coming from the equatorial channel. Recent investigations \citep{vol06,ldz06} have shown that the size and shape of the striped wind region, have important observational consequences. The region corresponding to the inner ring might split in several minor arches; the extent of the X-rays nebula changes as well as  the jet thickness. For the case of Crab Nebula simulations also seem to disfavor configuration with a higher magnetization close to the pole \citep{aro98}. While modeling of the equatorial region does not need any accurate treatment of the synchrotron losses (the feature are observed also in radio \citep{bie04}), these are fundamental in understanding the jet, and simulated map clearly show that the jet is not visible over the background, in optical or radio. 

Interestingly given that the particle energy distribution is somehow followed along a streamline, it is possible to reconstruct also spectral maps. In Fig.~3 a X-ray map of the spectral index based on simulations is presented. It reproduces the main observed properties of the Crab Nebula \citep{mor04}. We also see that simulations produce maps where the spectrum appears to flatten moving toward the main torus, without the need to assume any re-acceleration, this agree also with recent results about Vela \citep{pav07}. The reason is the Doppler boosting effect: for low speeds, observing at a given frequency implies sampling particle all at the same energy (if one neglects variations of magnetic field). When velocities are relativistic, the energy of particle responsible for emission at a given frequency, depends also on their Doppler boosting. The flatter spectrum in the torus, is due to the higher speed. There are still problems to recover the correct spectrum in the jet: the simulated spectrum is much steeper than what is observed, due to excessive synchrotron losses. To reproduce observations, one must assume some form of dissipation and re-energization along the axis. This could be associated with local instabilities of the toroidal magnetic field \citep{beg98} which are not captured by axisymmetric simulations, but for which there are many observational evidences \citep{pav03,mor04b,mel05,del06}. Integrated spectra can also be used to constrain the average nebular magnetization. Preliminary results show that $\sigma>0.03$ is required and configuration with a smaller striped wind region are favored. Interestingly integrated spectra show a flattening above the CHANDRA band, which has been interpreted as due to the high velocity in the posts-hock region. This could be a numerical artifact, or it may hint to the possibility of an injection break at high energies (either associated with acceleration process itself or with a changing efficiency along different portion of the TS). Further investigation will be necessary to understand this effect. 

%%%%%%%%%%%%%%%%%%%%%%%%%%%%%%%%%%% FIG 4 %%%%%%%%%%%%%%%%%%
\begin{figure}
\label{fig:4}
\resizebox{16cm}{!}{\includegraphics[bb=67 175 543 651, clip, scale=1.]{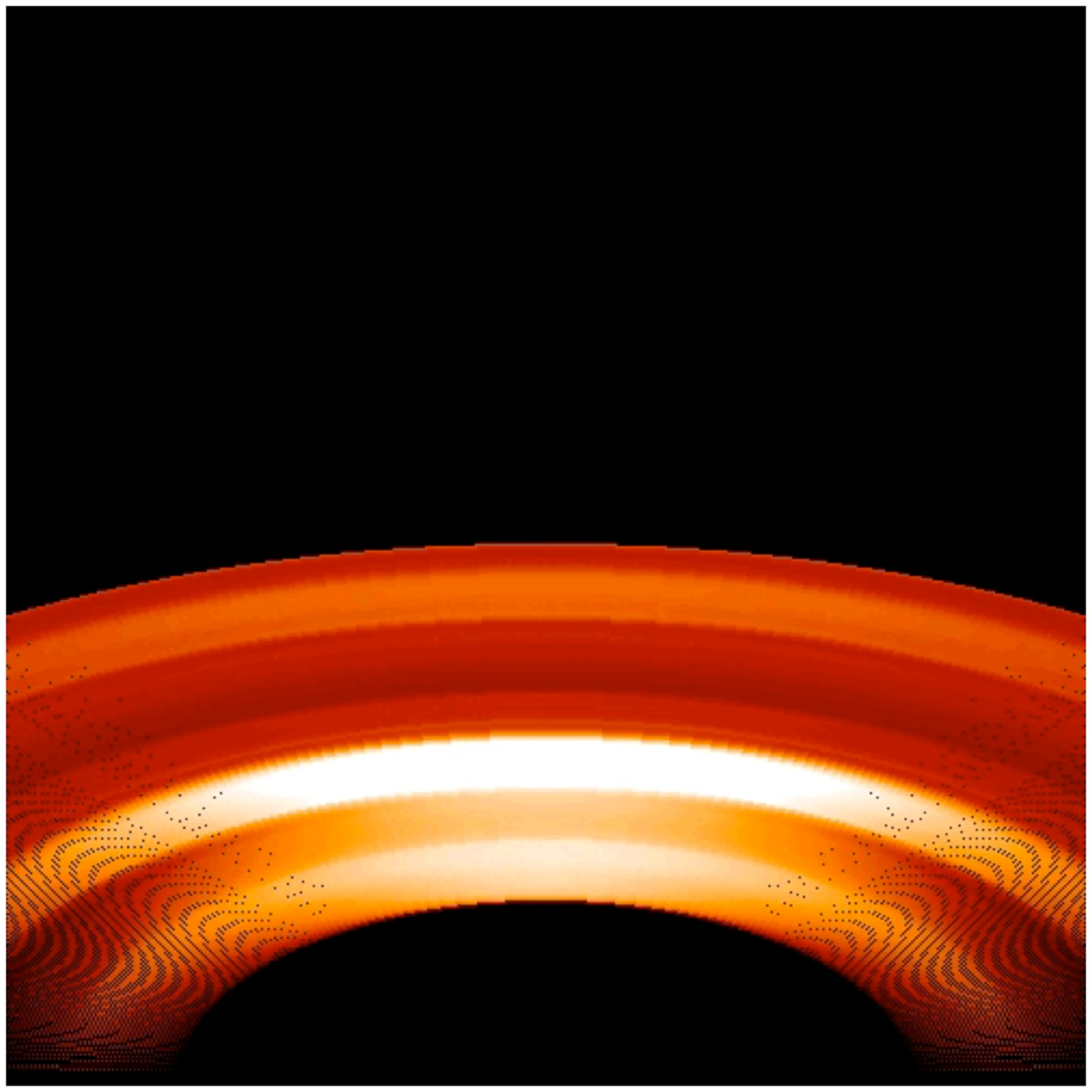}\hspace{0.5truecm}\includegraphics[bb=67 175 543 651, clip, scale=1.]{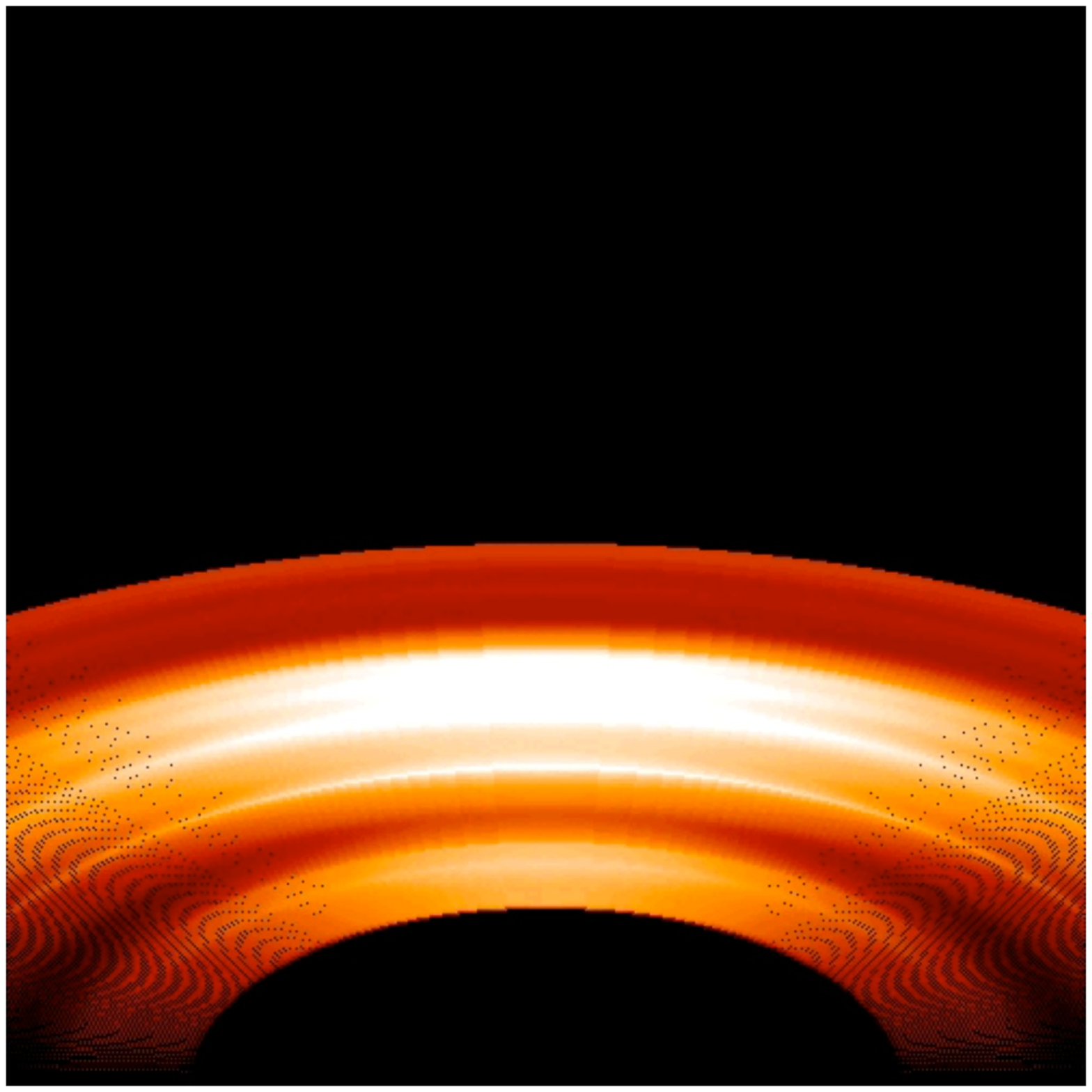}\hspace{0.5truecm}\includegraphics[bb=67 175 543 651, clip, scale=1.]{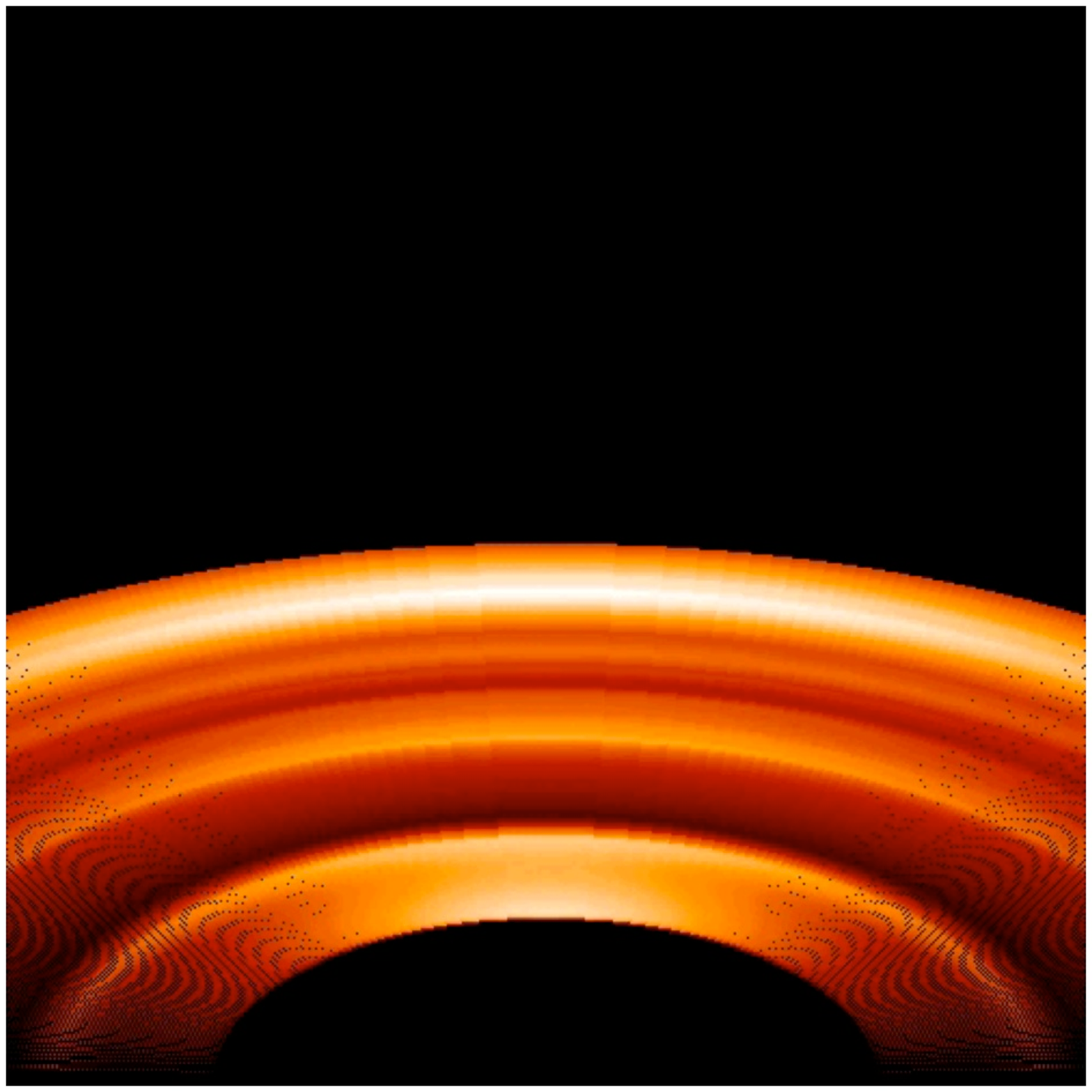}}
\caption{Variability in the wips region due to ion compression \citep{spi04}. The pictures represent the X-ray luminosity at three different instants. Notice the bright features moving outward.}
\end{figure}
%%%%%%%%%%%%%%%%%%%%%%%%%%%%%%%%%%%%%%%%%%%%%%%%%%%%%%%%%%%%

Polarization might also be used to derive useful information about the magnetic field structure. While emission maps are strongly dependent of the flow velocity inside the nebula, they are not much sensitive to the presence of a small scale disordered component of the magnetic field. In this case polarization might be  useful.  Given that only optical polarization is available, any study should be limited to the brightest features, like the wisps, inner ring and torus. The effect of flow velocity on the polarization angle has been discussed by \citet{me05b} and \citet{ldz06}. By comparison with old optical polarization map \citep{sch79}, one can immediately see that some of the properties are recovered, among whom a region which can be identified with the bended jet of Crab. More recent data (Graham, private communication) have also shown that the knot has a high degree of polarization, and that the polarization angle is consistent with the nebular origin of this feature. 

%%%%%%%%%%%%%%%%%%%%%%%%%%%%%%%%%%% FIG 5 %%%%%%%%%%%%%%%%%%
\begin{figure}
\label{fig:5}
\resizebox{7.8cm}{!}{
\includegraphics[scale=1]{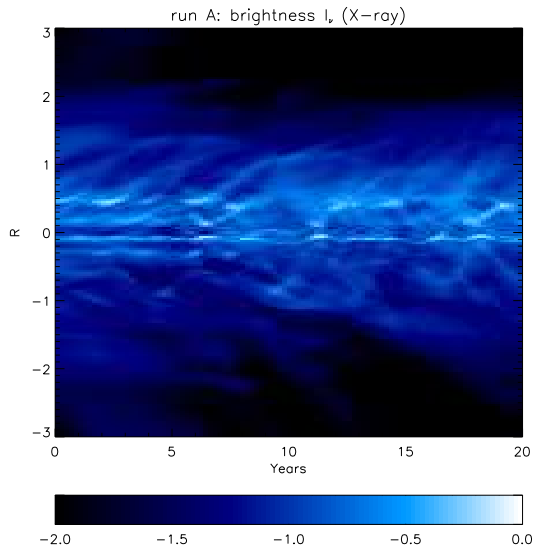}}
\caption{The Plot show the X-ray luminosity, based on MHD simulations, along the axis of a PWN as a function of time. Note the 2-year quasi periodicity in the wisps region and for the knot. Features appear to move out from the inner TS region. the propagation velocity is $\sim 0.5 c$ in the interior and slows down at larger distances.}
\end{figure}
%%%%%%%%%%%%%%%%%%%%%%%%%%%%%%%%%%%%%%%%%%%%%%%%%%%%%%%%%%%%
\section{Time variability}

The former discussion has focused on the explanation of the main observed features, that appear to be quite persistent on long time-scale. However it is known that PWNe show a short time variability at high energies. Variability of the wisps in the Crab Nebula has been known since the first high resolution observations \citep{hes02,bie04}. The jet in Vela appears to be strongly variables \citep{pav01,pav03}, and there is evidence for variability of the main rings \citep{pav07}. Variability is also observed in the jet of Crab \citep{mor04b,mel05}, and B1509 \citep{del06}. B1509 shows also variability in the inner ring. Variability in the jet, which usually has time-scale of years, has been associated to kink or sausage modes in the strong toroidal filed, or even to fire-hose instability \citep{tru88} in the case of Vela. On the other hand the wisps show variability on time-scale of months, much shorter than typical sound crossing time, in the form of an outgoing wave pattern, and recent results for the Crab Nebula suggest the presence of a year-long duty cycle.

moreFor a long time the only model capable of reproducing the observed variability was the one  proposed by \citet{spi04}. The model is based on the assumption that ions (or high energy electrons) are present in the equatorial region of the wind. The presence of ions is consistent with the fact that, in kinetic simulations of acceleration in a strong shock, a pure pair plasma do not produce any power-law distribution \citep{ama06}. Given that ions have a much larger larmor radius, of order of the size of the wisp region, they introduce a substantial deviation from a pure fluid picture: electrons are compresses by the gyration of ions and emission is enhanced. By selecting the fraction of energy in the ion component one can reproduce the observed time scale variability and the average distance between the torus and inner wisps. However results are based on a simplified 1D model that does not take into account the energy flux anisotropy, and the shape of the TS.

MHD nebular models however show the presence of high velocity flow channels inside the nebula, and it is possible that shear instabilities can be at the origin of the observed variability\citep{me06b,beg99}. It was already noted in simulations by \citet{bog05} that the  synchrotron emissivity inside the nebula varies. Results by \citet{kom04} also suggest that the flow might have a feedback action on the TS, causing it to change shape, and thus inducing a change in the wisps, which originate from posts-hock flow. It was not clear however if the observed variability in the internal flow pattern, could result in the observed variability of the wisp, if a wave pattern was recovered, and what were the time-scales. Recent results show that MHD models can account for the observed variability. The shear between the outgoing equatorial channel and the back-flow that is diverted by hoop stresses toward the axis, causes the formation of eddies on scales typical of the TS, that are subsequently advected away from the TS. These eddies in turn change the shape of the TS, and the Doppler boosting responsible for the arches and rings. Simulations show that there is a typical duty-cycle of about 1 year, and that the outgoing wave pattern is recovered. Typical speeds for such waves features is also in agreement with observations.

\section{Gamma rays}

Recently the interest of the scientific community has focused on the new gamma-ray results \citep{hof06,mal06}. HESS has been able in a few cases to resolve some PWNe both young ones in the free expansion phase and older one undergoing reverberation \citep{dej05,gal07}. New results will soon come with GLAST which will be able to investigate also the high energy MeV synchrotron emission from the tail of the particle energy distribution. Even if this results lack the high spatial resolution of the X-ray images they can provide useful information about PWNe.

The emission at MeV energy observed from young object will enable us to put constraints on the acceleration mechanism. Interestingly, given the variability of the wisps structure, one might expect that the emission at high energy should show similar variability, on comparable time-scales. On the other hand given the short lifetime for synchrotron losses of these particles, they might can be used to investigate short time variability in the wind (X-ray particles have longer lifetime). In principle, for example in the case of nulling pulsar \citep{kra07}, one could use the emission in this band to understand how the wind is modified in the null phase.

The emission at GeV energies is assumed to be from Inverse Compton scattering on background radiation, and in the case of young objects also the self synchrotron. Interestingly the spectral properties of the comptonized radiation can be used to derive information about particles which are supposed to emit synchrotron in the UV. Standard models for the evolution of the energy distribution function inside PWNe predict that the synchrotron break in the distribution should be in the region corresponding to these UV emitting particles. GeV emission could in principle help provide an independent constraint on  the magnetization in the nebula.  Moreover if protons are present in the pulsar wind, then it is possible that they might contribute to the gamma ray emission vis p-p scattering and pion decay \citep{ama03,hor06}.

Presently, evolutionary models for the high energy emission from PWNe, have been developed only in the case of young system during the free expansion. However, many of the PWN observed in the GeV band, are old objects in the reverberation phase, and it is important for a correct modeling of the emission to take into account the effect of the compression. This can lead to significant modifications both in the case of a pure leptonic model, due to the re-compression of the nebula, and also in the case of adronic emission, given that an efficient mixing with the colder SNR material is expected (thus increasing the density of target protons). 

In many of these objects the bulk of the gamma emission is not centered on the pulsar, and the displacement is too large to be explained in term of a moving pulsar leaving behind a relic PWN. Two possible explanations have been invoked: either a SNR exploding in a density gradient \citep{blo01}, or a bow-shock. In the first case the reverse shock will collapse  off-center with respect to the SNR, displacing the PWN. Observationally one should look for evidences of density gradients i he ISM, for example in the form of molecular clouds or star forming regions. In the second case, simulations and observations suggest that the pulsar wind back flow in the bow-shock tail can have high speeds up to fraction of the speed of light\citep{me05,kar07}. Particle responsible for the inverse compton emission in the gamma-ray band can be advected to large distances from the pulsar. In this case the displacement of the emission with respect to the pulsar should be aligned with the pulsar proper motion.

%%%%%%%%%%%%%%%%%%%%%%%%%%%%%%%%%%%%%%% SEC 5 %%%%%%%%%%%%%%%%%%%%%%%%%%%

\section{Conclusion}
\label{sec:concl}

In the last few years, the combination of high resolution observations, and numerical simulations, has improved our understanding of the evolution and internal dynamics of PWNe. We can reproduce the observed jet-torus structure and we can relate  the formation of the jet in the post shock flow to the wind magnetization. Simulated maps can reproduce many of the observed features, including the details of spectral properties. Results suggest that, the best agreement is achieved in the case of a striped wind, even if MHD simulations are not able to distinguish between dissipation of the current sheet in the wind or at the TS. Results also suggest that it is  possible to use X-rays imaging to constrain the pulsar wind properties; already the rings and tori observed in many PWNe have been used to determine the spin axis of the pulsar \citep{rom05}. Interestingly in Crab the inner ring does not appear boosted, while the wisps (which are interpreted as its optical counterpart) are.

There are still however unsolved questions, and possible future developments for research in this field.  All present simulations are axisymmetric, and none is able to address the problem of the stability of the toroidal field, and cannot reproduce the observed variability in the jet.  It is not clear if small scale disordered field is present in the inner region (may be a residual of the dissipation in the TS of the striped wind). A combination of simulations and polarimetry might help to answer this question.

%%%%%%%%%%%%%%%%%%%%%%%%%%%%%%%%%%%%%%%%%%%%%%%%
%% BACKMATTER
%%%%%%%%%%%%%%%%%%%%%%%%%%%%%%%%%%%%%%%%%%%%%%%%

\begin{theacknowledgments}
This work was supported partly by NASA through Hubble Fellowship grant HST-HF-01193.01-A, awarded by the Space Telescope Science Institute, which is operated by the Association of Universities for Research in Astronomy, Inc., for NASA, under contract NAS 5-26555. 
\end{theacknowledgments}

%% The code below assumes that BibTeX is used. Compliant BibTex styles
%% are aipproc (for use with natbib) and aipprocl (if natbib is missing
%% at the site).
%%
%% Please run "bibtex \jobname" to obtain the bibliography and 
%% then re-run LaTeX twice to fix the references!
%%
%% When referring to citations in the text, in quare brackets [] show
%% the number in order of appearance. References in the References
%% section are listed in the same numerical order.
%%%%%%%%%%%%%%%%%%%%%%%%%%%%%%%%%%%%%%%%%%%%%%%%%

\bibliographystyle{aipproc}   % if natbib is available


%\bibliography{sample}

\begin{thebibliography}{9}

\bibitem[Amato et al.(2003)]{ama03}
 Amato, E., Guetta, D., \& Blasi, P.\ 2003, \aap, 402, 827 

\bibitem[Amato \& Arons(2006)]{ama06}
 Amato, E., \& Arons, J.\ 2006, ArXiv Astrophysics e-prints, arXiv:astro-ph/0609034 

\bibitem[Arons(1998)]{aro98}
Arons, J.\ 1998, ``Neutron Star and Pulsar: Thirty years after Discovery'', Proc. International Conference on Neutron Stars and Pulsars, 17-20 Nov 1997, Rikko Univ., Tokyo

\bibitem[Bandiera et al.(1998)]{ban98}
 Bandiera, R., Amato, E., \& Woltjer, L.\ 1998, Memorie della Societa Astronomica Italiana, 69, 901 

\bibitem[Bandiera et al.(1999)]{ban99}
 Bandiera, R., Amato, E., Pacini, F., Salvati, M., \& Woltjer, L.\ 1999, Astrophysical Letters Communications, 38, 21 

\bibitem[Begelman(1998)]{beg98}
 Begelman, M.~C.\ 1998, \apj, 493, 291 

\bibitem[Begelman(1999)]{beg99}
 Begelman, M.~C.\ 1999, \apj, 512, 755 

\bibitem[Begelman \& Li(1992)]{beg92}
 Begelman, M.~C., \& Li, Z.-Y.\ 1992, \apj, 397, 187 

\bibitem[Begelman \& Li(1994)]{beg94} 
 Begelman, M.~C., \& Li, Z.-Y.\ 1994, ApJ, 426, 269

\bibitem[Beskin et al.(1998)]{bes98}
 Beskin, V.~S., Kuznetsova, I.~V., \& Rafikov, R.~R.\ 1998, \mnras, 299, 341 

\bibitem[Bietenholz et al.(1997)]{bie97}
 Bietenholz, M.~F., Kassim, N., Frail, D.~A., Perley, R.~A., Erickson, W.~C., \& Hajian, A.~R.\ 1997, \apj, 490, 291 

\bibitem[Bietenholz et al.(2004)]{bie04}
 Bietenholz, M.~F., Hester, J.~J., Frail, D.~A., \& Bartel, N.\ 2004, \apj, 615, 794 

\bibitem[Blondin et al.(2001)]{blo01}
 Blondin, J.~M., Chevalier, R.~A., \& Frierson, D.~M.\ 2001, \apj, 563, 806 

\bibitem[Bogovalov(2001a)]{bog01}
 Bogovalov, S.~V.\ 2001a, \aap, 371, 1155 

\bibitem[Bogovalov(2001b)]{bog01b}
 Bogovalov, S.~V.\ 2001b, \aap, 367, 159 

\bibitem[Bogovalov \& Khangoulian(2002a)]{bog02a}
 Bogovalov, S.~V., \& Khangoulian, D.~V.\ 2002a, \mnras, 336, L53 

\bibitem[Bogovalov \& Khangoulyan(2002b)]{bog02b}\
 Bogovalov, S.~V., \& Khangoulyan, D.~V.\ 2002b, Astronomy Letters, 28, 373 

\bibitem[Bogovalov et al.(2005)]{bog05}
 Bogovalov, S.~V., Chechetkin, V.~M., Koldoba, A.~V., \& Ustyugova, G.~V.\ 2005, \mnras, 358, 705

\bibitem[Bucciantini \& Bandiera(2001)]{me01}
 Bucciantini, N., \& Bandiera, R.\ 2001, \aap, 375, 1032 

\bibitem[Bucciantini(2002)]{me02}
 Bucciantini, N.\ 2002, \aap, 387, 1066  

\bibitem[Bucciantini et al.(2003)]{me03}
 Bucciantini, N., Blondin, J.~M., Del Zanna, L., \& Amato, E.\ 2003, \aap, 405, 617 

\bibitem[Bucciantini et al.(2004a)]{me04}
 Bucciantini, N., Bandiera, R., Blondin, J.~M., Amato, E., \& Del Zanna, L.\ 2004a, \aap, 422, 609 

\bibitem[Bucciantini et al.(2004)]{me04b}
 Bucciantini, N., Amato, E., Bandiera, R., Blondin, J.~M., \& Del Zanna, L.\ 2004b, \aap, 423, 253 

\bibitem[Bucciantini et al.(2005a)]{me05}
 Bucciantini, N., Amato, E., \& Del Zanna, L.\ 2005a, \aap, 434, 189 

\bibitem[Bucciantini et al.(2005b)]{me05b} Bucciantini, N., 
del Zanna, L., Amato, E., \& Volpi, D.\ 2005b, \aap, 443, 519 

\bibitem[Bucciantini et al.(2006a)]{me06}
 Bucciantini, N., Thompson, T.~A., Arons, J., Quataert, E., \& Del Zanna, L.\ 2006a, \mnras, 368, 1717 

\bibitem[Bucciantini \& Del Zanna(2006b)]{me06b}
 Bucciantini, N., \& Del Zanna, L.\ 2006b, \aap, 454, 393 

\bibitem[Contopoulos et al.(1999)]{con99}
 Contopoulos, I., Kazanas, D., \& Fendt, C.\ 1999, \apj, 511, 351 

\bibitem[Camilo et al.(2004)]{cam04}
 Camilo, F., Gaensler, B.~M., Gotthelf, E.~V., Halpern, J.~P., \& Manchester, R.~N.\ 2004, \apj, 616, 1118 

\bibitem[de Jager(2005)]{dej05}
de Jager, O.~C.\ 2005, Astrophysical Sources of High Energy Particles and Radiation, 801, 298 

\bibitem[DeLaney et al.(2006)]{del06}
 DeLaney, T., Gaensler, B.~M., Arons, J., \& Pivovaroff, M.~J.\ 2006, \apj, 640, 929 

\bibitem[Del Zanna et al.(2003)]{ldz03}
 Del Zanna, L., Bucciantini, N., \& Londrillo, P.\ 2003, \aap, 400, 397 

\bibitem[Del Zanna et al.(2004)]{ldz04}
 Del Zanna, L., Amato, E., \& Bucciantini, N.\ 2004, \aap, 421, 1063 

\bibitem[Del Zanna et al.(2006)]{ldz06}
 Del Zanna, L., Volpi, D., Amato, E., \& Bucciantini, N.\ 2006, \aap, 453, 621 

\bibitem[Gaensler et al.(2002)]{gae02}
 Gaensler, B.~M., Arons, J., Kaspi, V.~M., Pivovaroff, M.~J., Kawai, N., \& Tamura, K.\ 2002, \apj, 569, 878 

\bibitem[Gaensler et al.(2001)]{gae01}
 Gaensler, B.~M., Pivovaroff, M.~J., \& Garmire, G.~P.\ 2001, \apjl, 556, L107 

\bibitem[Gaensler \& Slane(2006)]{gae06}
 Gaensler, B.~M., \& Slane, P.~O.\ 2006, ARA\&A, 44, 17 

\bibitem[Gallant et al. (2007)]{gal07}
 Gallant, Y.~A., \& al.\ 2007, ``VHE gamma-ray emitting pulsar wind nebulae discovered by HESS'', these proceedings

\bibitem[Gammie et al.(2003)]{gam03}
 Gammie, C.~F., McKinney, J.~C., \& T{\'o}th, G.\ 2003, \apj, 589, 444 

\bibitem[Gotthelf \& Wang(2000)]{got00}
 Gotthelf, E.~V., \& Wang, Q.~D.\ 2000, \apjl, 532, L117 

\bibitem[Gruzinov(2005)]{gru05}
 Gruzinov, A.\ 2005, Physical Review Letters, 94, 021101 

\bibitem[Helfand et al.(2001)]{hel01}
 Helfand, D.~J., Gotthelf, E.~V., \& Halpern, J.~P.\ 2001, \apj, 556, 380 

\bibitem[Hester et al.(1995)]{hes95}
 Hester, J.~J., et al.\ 1995, \apj, 448, 240 

\bibitem[Hester et al.(1996)]{hes96}
 Hester, J.~J., et al.\ 1996, \apj, 456, 225 

\bibitem[Hester et al.(2002)]{hes02}
 Hester, J.~J., et al.\ 2002, \apjl, 577, L49 

\bibitem[Hickson \& van den Bergh(1990)]{hic90}
 Hickson, P., \& van den Bergh, S.\ 1990, \apj, 365, 224 

\bibitem[Hoffmann et al.(2007)]{hof06}
 Hoffmann, A.~I.~D., Horns, D., \& Santangelo, A.\ 2007, A\&SS, 309, 215 

\bibitem[Horns et al.(2006)]{hor06}
 Horns, D., Aharonian, F., Santangelo, A., Hoffmann, A.~I.~D., \& Masterson, C.\ 2006, \aap, 451, L51 

\bibitem[Jun(1998)]{jun98}
 Jun, B.-I.\ 1998, \apj, 499, 282 

\bibitem[Kargaltsev et al.(2007)]{kar07}
 Kargaltsev, O., Pavolv, G.~G., Misanivic, Z., \& Garmire, G.~P.\ 2007, ``X-ray observation of pulsar tails'', these proceedings. 

\bibitem[Khangoulian \& Bogovalov(2003)]{kan03}
 Khangoulian, D.~V., \& Bogovalov, S.~V.\ 2003, Astronomy Letters, 29, 495 

\bibitem[Kennel \& Coroniti(1984a)]{ken84a}
 Kennel, C.~F., \& Coroniti, F.~V.\ 1984a, \apj, 283, 710, KC84 

\bibitem[Kennel \& Coroniti(1984b)]{ken84b}
 Kennel, C.~F., \& Coroniti, F.~V.\ 1984b, \apj, 283, 694, KC84 

\bibitem[Kirk \& Skj{\ae}raasen(2003)]{kir03}
 Kirk, J.~G., \& Skj{\ae}raasen, O.\ 2003, \apj, 591, 366 

\bibitem[Komissarov(1999)]{kom99}
 Komissarov, S.~S.\ 1999, \mnras, 303, 343 

\bibitem[Komissarov(2006)]{kom06}
 Komissarov, S.~S.\ 2006, \mnras, 367, 19 

\bibitem[Komissarov \& Lyubarsky(2004)]{kom04}
 Komissarov, S.~S., \& Lyubarsky, Y.~E.\ 2004, \mnras, 349, 779

\bibitem[Kramer(2007)]{kra07}
 Kramer, M.\ 2007, ``Observation of pulsed emission from pulsars'', these proceedings 

\bibitem[Melatos et al.(2005)]{mel05}
 Melatos, A., et al.\ 2005, \apj, 633, 931 

 \bibitem[Lu et al.(2002)]{lu02}
 Lu, F.~J., Wang, Q.~D., Aschenbach, B., Durouchoux, P., \& Song, L.~M.\ 2002, \apjl, 568, L49 

\bibitem[Lyubarsky(2002)]{lyu02}
 Lyubarsky, Y.~E.\ 2002, \mnras, 329, L34 

\bibitem[Lyubarsky(2003)]{lyu03b}
 Lyubarsky, Y.~E.\ 2003, \mnras, 345, 153

\bibitem[Lyubarsky(2005)]{lyu05}
 Lyubarsky, Y.\ 2005, Advances in Space Research, 35, 1112 

\bibitem[Lyubarsky \& Eichler(2001)]{lyu01}
 Lyubarsky, Y., \& Eichler, D.\ 2001, \apj, 562, 494 

\bibitem[Lyubarsky \& Kirk(2001)]{lyu01b}
 Lyubarsky, Y., \& Kirk, J.~G.\ 2001, \apj, 547, 437 

\bibitem[Lyutikov(2003)]{lyu03}
 Lyutikov, M.\ 2003, \mnras, 339, 623 

\bibitem[Michel(1973)]{mic73}
 Michel, F.~C.\ 1973, \apjl, 180, L133 

\bibitem[Michel et al.(1991)]{mic91}
 Michel, F.~C., Scowen, P.~A., Dufour, R.~J., \& Hester, J.~J.\ 1991, \apj, 368, 463 

\bibitem[Mori et al.(2004)]{mor04}
 Mori, K., Burrows, D.~N., Hester, J.~J., Pavlov, G.~G., Shibata, S., \& Tsunemi, H.\ 2004, \apj, 609, 186 

\bibitem[Mori et al.(2004)]{mor04b}
 Mori, K., Burrows, D.~N., Pavlov, G.~G., Hester, J.~J., Shibata, S., \& Tsunemi, H.\ 2004b, IAU Symposium, 218, 181 

\bibitem[Pavlov et al.(2001)]{pav01}
 Pavlov, G.~G., Kargaltsev, O.~Y., Sanwal, D., \& Garmire, G.~P.\ 2001, \apjl, 554, L189 

\bibitem[Pavlov et al.(2003)]{pav03}
 Pavlov, G.~G., Teter, M.~A., Kargaltsev, O., \& Sanwal, D.\ 2003, \apj, 591, 1157 

\bibitem[Pavlov (2007)]{pav07}
 Pavlov, G.~G.\ 2007, ``Pulsar wind nebulae in the CHANDRA era'', these proceedings

\bibitem[Rees \& Gunn(1974)]{ree74}
 Rees, M.~J., \& Gunn, J.~E.\ 1974, \mnras, 167, 1 

\bibitem[Reynolds \& Chevalier(1984)]{ren84}
 Reynolds, S.~P., \& Chevalier, R.~A.\ 1984, \apj, 278, 630 

\bibitem[Roberts et al.(2006)]{mal06}
 Roberts, M.~S.~E., Gotthelf, E.~V., Halpern, J.~P., Brogan, C.~L., \& Ransom, S.~M.\ 2006, arXiv:astro-ph/0612631 

\bibitem[Romani \& Ng(2003)]{rom03}
 Romani, R.~W., \& Ng, C.-Y.\ 2003, \apjl, 585, L41 

\bibitem[Romani et al.(2005)]{rom05}
 Romani, R.~W., Ng, C.-Y., Dodson, R., \& Brisken, W.\ 2005, \apj, 631, 480 

\bibitem[Schmidt et al.(1979)]{sch79}
 Schmidt, G.~D., Angel, J.~R.~P., \& Beaver, E.~A.\ 1979, \apj, 227, 106 

\bibitem[Shibata et al.(2003)]{shi03}
 Shibata, S., Tomatsuri, H., Shimanuki, M., Saito, K., \& Mori, K.\ 2003, \mnras, 346, 841 

\bibitem[Slane et al.(2004)]{sla04}
 Slane, P., Helfand, D.~J., van der Swaluw, E., \& Murray, S.~S.\ 2004, \apj, 616, 403 

\bibitem[Spitkovsky(2006)]{spi06}
 Spitkovsky, A.\ 2006, \apjl, 648, L51 

\bibitem[Spitkovsky \& Arons(2004)]{spi04}
 Spitkovsky, A., \& Arons, J.\ 2004, \apj, 603, 669 

\bibitem[Trussoni et al.(1988)]{tru88}
 Trussoni, E., Massaglia, S., Bodo, G., \& Ferrari, A.\ 1988, \mnras, 234, 539 

\bibitem[van der Swaluw(2003)]{van03}
 van der Swaluw, E.\ 2003, \aap, 404, 939 

\bibitem[van der Swaluw et al.(2001)]{van01}
 van der Swaluw, E., Achterberg, A., Gallant, Y.~A., \& T{\'o}th, G.\ 2001, \aap, 380, 309 

\bibitem[van der Swaluw et al.(2003)]{van03b}
 van der Swaluw, E., Achterberg, A., Gallant, Y.~A., Downes, T.~P., \& Keppens, R.\ 2003, \aap, 397, 913 

\bibitem[van der Swaluw et al.(2004)]{van04}
 van der Swaluw, E., Downes, T.~P., \& Keegan, R.\ 2004, \aap, 420, 937 

\bibitem[Veron-Cetty \& Woltjer(1993)]{ver93}
 Veron-Cetty, M.~P., \& Woltjer, L.\ 1993, \aap, 270, 370 

\bibitem[Velusamy(1985)]{vel85}
 Velusamy, T.\ 1985, \mnras, 212, 359 

\bibitem[Volpi(2006)]{vol06}
 Volpi, D./ 2006, Thesis: ``La Strutura Interna della Nebulosa del Granchio''

\bibitem[Weisskopf et al.(2000)]{wei00}
 Weisskopf, M.~C., et al.\ 2000, \apjl, 536, L81 

\bibitem[Willingale et al.(2001)]{wil01}
 Willingale, R., Aschenbach, B., Griffiths, R.~G., Sembay, S., Warwick, R.~S., Becker, W., Abbey, A.~F., \& Bonnet-Bidaud, J.-M.\ 2001, \aap, 365, L212 

\bibitem[Wilson(1972)]{wil72}
 Wilson, A.~S.\ 1972, \mnras, 157, 229 




\end{thebibliography}

%%%%%%%%%%%%%%%%%%%%%%%%%%%%%%%%%%%%%%%%%%%%%%%%%
%% If the bibliography is
%% produced without BibTeX, comment out the above lines, use
%% \begin{thebibliography}{widest-label} environment to hold 
%% the list of references and 
%% \bibitem{label} command to start a bibliographical entry having
%% the "label" for use in \cite commands.

%\end{document}

%%%%%%%%%%%%%%%%%%%%%%%%%%%%%%%%%%%%%%%%%%%
%% The following lines show an example how to produce a bibliography
%% without the help of the BibTeX program. This could be used instead
%% of the above.
%%%%%%%%%%%%%%%%%%%%%%%%%%%%%%%%%%%%%%%%%%%

\end{document}